\documentclass[sigconf]{acmart}

\usepackage{booktabs}

\newcommand{\RQOne}{Do development teams share the same affective states?}
\newcommand{\RQTwo}{Can affective states predict today's productivity and social interaction?}
\newcommand{\RQThree}{Can today's productivity and social interaction predict tomorrow's affective states?} 

\begin{document}
\copyrightyear{2018} 
\acmYear{2018} 
\setcopyright{acmlicensed}
\acmConference[ESEM '18]{ACM / IEEE International Symposium on Empirical Software Engineering and Measurement (ESEM)}{October 11--12, 2018}{Oulu, Finland}
\acmBooktitle{ACM / IEEE International Symposium on Empirical Software Engineering and Measurement (ESEM) (ESEM '18), October 11--12, 2018, Oulu, Finland}
\acmPrice{15.00}
\acmDOI{10.1145/3239235.3239245}
\acmISBN{978-1-4503-5823-1/18/10}

\title{Using Experience Sampling to link Software Repositories with Emotions and Work Well-Being}

\author{Miikka Kuutila}
\affiliation{M3S, ITEE, University of Oulu, Finland}
\email{miikka.kuutila@oulu.fi}

\author{Mika V. M{\"a}ntyl{\"a}}
\affiliation{M3S, ITEE, University of Oulu, Finland}
\email{mika.mantyla@oulu.fi}

\author{Ma{\"e}lick Claes}
\affiliation{M3S, ITEE, University of Oulu, Finland}
\email{maelick.claes@oulu.fi}

\author{Marko Elovainio}
\affiliation{Department of Psychology and Logopedics, University of Helsinki, Finland}
\email{marko.elovainio@helsinki.fi}

\author{Bram Adams}
\affiliation{MCIS, Polytechnique Montreal, Canada}
\email{bram.adams@polymtl.ca}

\renewcommand{\shortauthors}{M.Kuutila et al.}
\renewcommand{\shorttitle}{Using Experience Sampling to link Software Repositories}

\begin{abstract}
\textbf{Background:} The experience sampling method studies everyday experiences of humans in natural environments. In psychology it has been used to study the relationships between work well-being and  productivity. To our best knowledge, daily experience sampling has not been previously used in software engineering.
\textbf{Aims:} Our aim is to identify links between software developers self-reported affective states and work well-being and measures obtained from software repositories. 
\textbf{Method:} We perform an experience sampling study in a software company for a period of eight months, we use logistic regression to link the well-being measures with development activities, i.e. number of commits and chat messages.
\textbf{Results:} We find several significant relationships between questionnaire variables and software repository variables. To our surprise relationship between hurry and number of commits is negative, meaning more perceived hurry is linked with a smaller number of commits. We also find a negative relationship between social interaction and hindered work well-being. 
\textbf{Conclusions:} The negative link between commits and hurry is counter-intuitive and goes against previous lab-experiments in software engineering that show increased efficiency under time pressure. Overall, our work is an initial step in using experience sampling in software engineering and validating theories on work well-being from other fields in the domain of software engineering.
\end{abstract}

\ccsdesc[100]{Software and its engineering~Programming teams}

\keywords{Experience Sampling, Repository mining, binary regression, affective states, empirical study, work well-being, stress, time pressure}

\maketitle
\section{Introduction}\label{sec:introduction}

In recent years, the affective and cognitive states of software developers have drawn more interest of study by academics. People have called for more work on the topics such as ``behavioral software engineering'' \cite{lenberg2015behavioral}, that borrows its name from the field of behavioral economics, and ``psychoempirical software engineering'' \cite{graziotin2015understanding}. Software engineering researchers have established focused venues to study the affective state of software developers such as \textit{The International Workshop on Emotion Awareness in Software Engineering}\footnote{http://collab.di.uniba.it/semotion/}

Past work in this area has taken many forms. Controlled experiments in the lab have investigated the impact of sleep deprivation~\cite{fucci2018need}, emotional arousal induced by time pressure~\cite{mantyla2014time} and emotional states of dominance and valence~\cite{graziotin2013happy}. Quantitative~\cite{nan2009impact} and qualitative studies~\cite{linssen2018antecedents} have showed the impact of time pressure on the field. Studies on mining software repositories have made several recent attempts to build tools and reason about the emotional states of software developers by utilizing sentiment analysis ~\cite{mantyla2017bootstrapping,lin2018sentiment,novielli2018benchmark, jongeling2017negative,islam2017leveraging}. However, to our best knowledge no prior studies have attempted to link daily self-reported experiences of emotions or work well-being with measures from software repositories. 

We investigate if different stressors are felt on an individual or organizational level, and try to establish links between software development activities and self-reported feelings of well-being. To achieve our goal, we used experience sampling methodology and created a questionnaire to be taken daily in an industrial software project setting. This questionnaire assesses individual developers self-reported hurry, stress, sleeping problems, interruptions, ineffective software development (defined as poorly working tools, poor processes or communication) and independence. Metrics obtained with the questionnaire are then linked to the metrics of productivity and social interaction obtained from software repositories with logistic regression models. Our research questions are formulated as:

\begin{description}
\item[RQ1] {\RQOne}
\item[RQ2] {\RQTwo}
\item[RQ3] {\RQThree}
\end{description}

The rest of the paper is structured as follows. The experience sampling methodology from the field of psychology and some relevant prior work is introduced in Section~\ref{sec:experience}. The methodology for creating the daily questionnaire and executing this study is explained in Section~\ref{sec:methodology}. In Section~\ref{sec:results} we provide motivation for the investigation for our research questions, present the results and discuss them. We discuss threats to validity and future work in Section~\ref{sec:future}. Lastly, conclusions are provided in Section~\ref{sec:conclusions}.
\section{Experience Sampling Method}\label{sec:experience}
\subsection{Overview from Psychology}
Experience sampling method (ESM), also know as the Daily Diary method, studies everyday experiences and behavior in natural environment, with data gathered both from psychological and physiological sources ~\cite{alliger1993using}. Strengths of experience sampling methodology are its empirical nature in the documentation of real-life experiences increasing its ecological validity, its allowance of investigating within-person processes, its reduction of memory bias compared with other methods using self-reports, its allowance of investigating contingent behavior, and lastly its ability to augment other research methods. Among possible weaknesses related to experience sampling are the self-selection bias, motivation issues in the acquired sample, the limited number of questions in data gathering, and the possible reactivity to research setting ~\cite{scollon2009experience}.

Experience sampling methods have been divided into three categories~\cite{scollon2009experience} based on the time when the experiences are gathered: Interval-contingent sampling, event-contingent sampling and signal-contingent sampling. Interval-contingent refers to collecting the data after a given time interval (e.g. hourly, daily or weekly). In event-contingent data is gathered after specific events (e.g. after every meeting or social interaction). Lastly signal-contingent refers to a situation where where participants in the study are prompted to answer at a randomly timed signal. A variety of devices can be used to remind subjects to respond to surveys and questionnaires, such as personal digital assistants, booklets, beepers or wristwatches ~\cite{kimhy2006computerized}. However, reminders with email or SMS are also commonly applied. 

In prior studies on work well-being, experience sampling methods and daily questionnaires have been used to study events, moods and behavior in work setting. Some examples of findings are that negative job events are five times more related to negative mood than positive job events to positive mood ~\cite{miner2005experience}. Additionally, job satisfaction has been measured with experience sampling methodology and evidence has been found for that affect and cognition are antecedents to job satisfaction ~\cite{ilies2004experience}. This motivates our paper title as we think experienced work well-being and experienced emotions cannot really be separated.  Continued cognitive engagement, more positive affect during work activity than during leisure activity, and preference of work activities over leisure activities have been linked to workaholism in an ESM study ~\cite{snir2008workaholism}. Outside work context Experience sampling has also been used to study interaction with information systems, for example a study found out at that increase in the usage of Facebook predicted lower life satisfaction level~\cite{kross2013facebook}.

\subsection{Challenges in Statistical Analysis}
Experience sampling methods produce time-series data that should not be analyzed by typical statistical methods.  
West and Hepworth~\cite{west1991statistical} discuss the statistical issues of analyzing the data of daily experiences. As statistical tests assume the independence of observations, non-independence in the time series data gathered with experience sampling is a problem needing action. West and Hepworth~\cite{west1991statistical} identify three main sources of non-independence which can occur in the data: autocorrelation, trend, and seasonality and which should be accounted in the analysis. 

Repeated measures over time can create autocorrelation, i.e. time dependent data in violation of the assumption of independence. For example, level of stress felt today is not completely independent on the level of stress felt yesterday. Controlling for trend is important when cross-correlating time series, as underlying trends create spurious correlations between the time series. For example, increasing trend in the number of software engineers over time would create spurious correlations with many software engineering output measures such as commits and defect reports.  Seasonality components usually refer to weekly, monthly or yearly cycles, for example stress could be felt more on Mondays.

\section{Methodology}\label{sec:methodology}

Our experience sampling study was conducted in a medium-sized software company in Finland. We developed a questionnaire that we sent to a team developing a service with Agile methods and Continuous Delivery. The project has a single customer and meetings with the customer are held almost weekly. The project was originally started in 2014. Additionally we gathered data about the development activity from the Git repository and the chat system used internally by the company.

\subsection{Daily questionnaire}

We constructed a small questionnaire ~\cite{kuutila2018daily}, to be taken daily at the software company, with the goal of producing data related to work experiences of the software project personnel. We piloted the questionnaire with the authors, to achieve our goal of producing a questionnaire which can be taken quickly to achieve high response rates. Single item measurements have been shown to produce valid data in prior studies ~\cite{wanous1997overall, nagy2002using, elo2003validity}. The questionnaire includes six single items that measure variables related to job well-being on a Likert-scale. The questionnaire was constructed by picking relevant items on the survey done by Heponiemi et al.~\cite{heponiemi2017finnish}. The fourth author of this paper had participated in the study  ~\cite{heponiemi2017finnish} and has extensive experience in utilizing work health questionnaires in multiple domains. Thus, our questions represent well understood theoretical concepts in work health. As the past survey ~\cite{heponiemi2017finnish} was not done in software engineering we added one software engineering specific item. To make sure respondents can answer quickly, we decided to include only one software specific question in the questionnaire:
\begin{itemize}
\item I can make independent decisions in my work 
\item I am in a hurry and have too little time to finish the task properly
\item I feel interrupted while working
\item I experience ineffective software development (poor processes, poorly performing tools or poor communication with the development team)
\item I feel stressed (refers to a situation in which the respondent feels tense, restless, nervous or anxious)
\item I experience sleeping problems (difficulty in falling asleep or waking up several times during the night)
\end{itemize}

One definition of \textbf{stress} is a relationship between individual and environment, that is taxing or  exceeding resources for coping ~\cite{folkman2013stress}. In job demands-resources model~\cite{bakker2007job}, stress is produced by the imbalance between job resources and demands. Individuals \textbf{independence} and autonomy have been under study as a mediating factor between job demands and resources ~\cite{bakker2005job,xanthopoulou2007job}, i.e. there is evidence that increased autonomy in work tasks lessens the effects job demands such as time pressure. \textbf{Hurry} to complete work, also known as time pressure, is a job demand, and has been shown to be associated with increased performance~\cite{nan2009impact,mantyla2014time}, but also higher stress~\cite{svenson1993time} and even burnout~\cite{donald2005work,bakker2005job,sonnentag1994stressor}. \textbf{Sleeping problems} have been strongly linked to stress and increased job demands ~\cite{aakerstedt2002sleep, linton2004does}. \textbf{Interruptions} to work increase effort needed for task completion and have also been shown to increase time pressure and stress in software development context ~\cite{mark2008cost}. The last item we included in our survey (ineffective software development) was software development specific, which included common topics related to productivity in software processes~\cite{diaz1997software}, tools\cite{bruckhaus1996impact} and communication\cite{wagner2018systematic}. 

The respondents were asked to rate these six items with the question: "How frequently has the following condition occurred since the last time you answered this survey?". These items were then ranked on a five-point Likert-scale. From 1 to 5, the corresponding textual answers are "Very rarely or never", "Rarely", "Once in a while", "Often" and "Frequently or continuously". Before starting the data collection, we met with the project personnel to explain the purpose of the study, the purpose of capturing frequency as well as intensity with the scale of the questionnaire and the voluntary nature of participation to the study.

The developed questionnaire was sent to the developers of the project from April 10th 2017 to January 12th 2018. We used Webropol\footnote{\url{http://w3.webropol.com/start/}} to send the questionnaire every working day by email at 8am and to collect the responses. Developers who moved from or to another project, or started working in multiple projects at the same time, stopped answering the questionnaire. Developers with less than ten responses were discarded from the data analysis. For data analysis, a total of 526 responses were received from eight respondents. Multiple answers received during the same day by one individual were annexed with the mean of those answers, reducing the amount of analyzable answers to 502. Taking into account summer holidays, the total response rate is 37,5\% (526 / 1404) for eligible respondents.

\subsection{Mining Software Repositories}

After the gathering of the experience sampling data with the developed questionnaire, we extracted data from the available software repositories of the project. We used GrimoireLab\footnote{\url{https://grimoirelab.github.io/}} to extract the list of commits from the Git repository used by the project team. For each day of the period during which the developers answered the questionnaire, we computed for each respondent the number of commits made (\emph{ncommits}) and the number of lines of code changes (\emph{nloc}). While software development contains tasks not captured by these metrics, number of commits and lines of code have been widely used as proxy measures for productivity in prior literature. E.g. commits in seminal work in software related case studies ~\cite{mockus2002two} and lines of code in cost estimation models since at least the late seventies ~\cite{boehm1981software}. 

Additionally, the company also provided us with a JSON dump of the chat room used by the developers. From this chat archive, we computed the daily number of chat messages (\emph{nchat}) for each respondent. The specific tool used for communication changed during our study, hipchat \footnote{\url{http://www.businessinsider.com/atlassian-launches-hipchat-successor-stride-2017-9}} and slack \footnote{\url{https://slack.com/}} were used.

\subsection{Analysis}

\begin{figure*}[!t]
\caption{Outcome of decompose-function for number of chat messages without weekends.}
\includegraphics[scale=0.50]{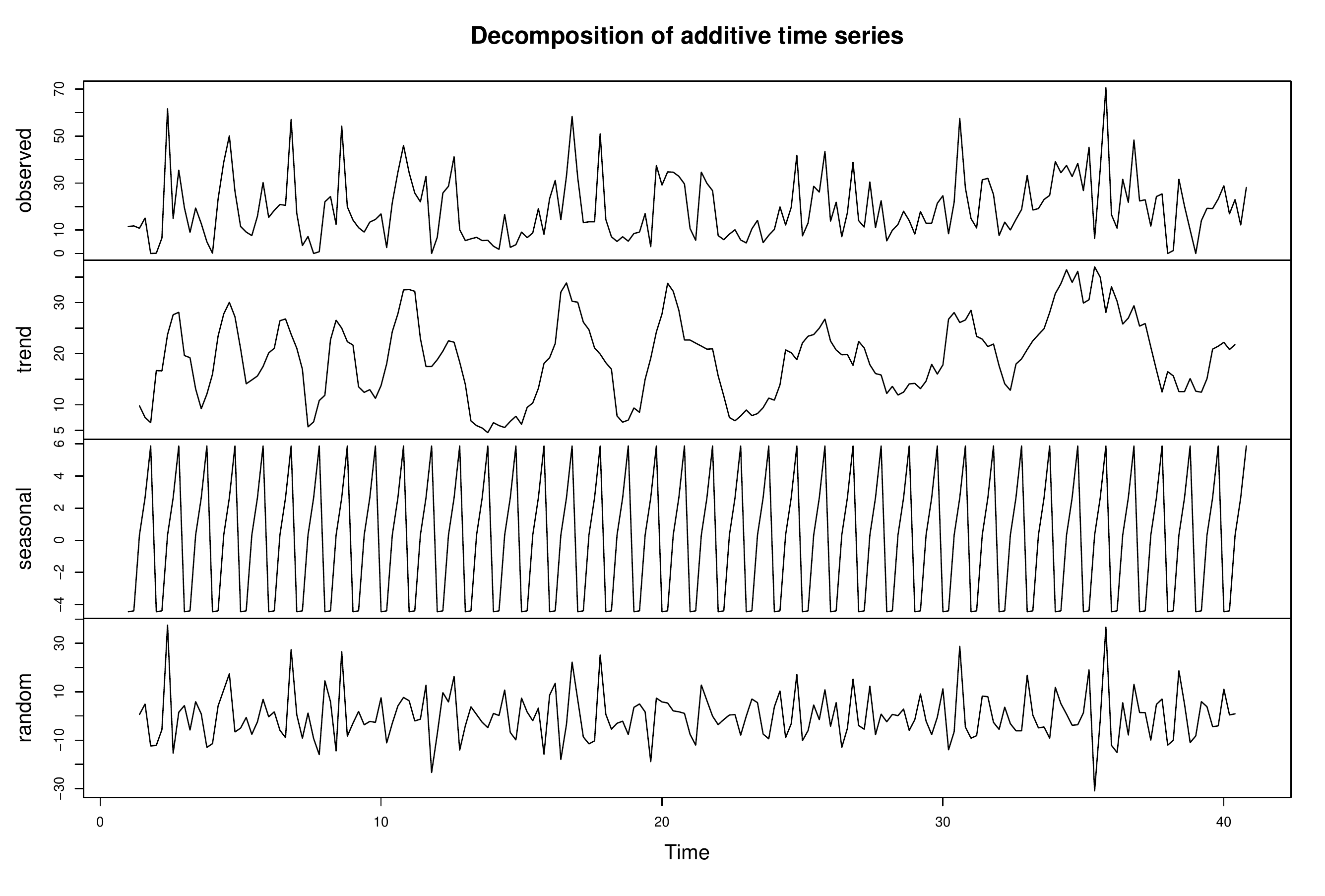}
\label{fig:seasonalityOfChat}
\end{figure*}

We studied the trends and seasonality in our collected data with the R function \emph{decompose}\footnote{\url{https://www.rdocumentation.org/packages/stats/versions/3.4.3/topics/decompose}} and found small weekly seasonality for all the software repository variables. Weekly seasonality of the chat messages is the highest, with ten messages being the difference between most active and least active days for chat. By comparison, the seasonality of commits per day is much weaker, with an average difference of 0.4 commits between the most active and least active day of the week. Figure \ref{fig:seasonalityOfChat} shows a graph produced by the decompose function and the seasonality of chat messages. To account for time-series data and control for weekly seasonality in the data, we added dummy boolean variables for each day to all regression models; e.g. if dummy variable Monday is True all other dummy variables for weekdays are False. 

We also investigated the autocorrelations of the data with the \emph{acf} function\footnote{\url{https://www.rdocumentation.org/packages/forecast/versions/8.3/topics/Acf}} of the forecast R package~\cite{hyndman2007automatic}. We found significant autocorrelations for all of the questionnaire variables. As a consequence, we entered a lagged variable for all questionnaire variables when they were used as input variables to the equations in our regression models, i.e. the variable itself of the previous day (t-1) when using the same variable, the day (t) as the outcome variable. This has been done in a variety of experience sampling studies to prevent autocorrelation, as advocated by West and Hepworth~\cite{west1991statistical}.

We used base R's cut-function\footnote{\url{https://www.rdocumentation.org/packages/base/versions/3.5.0/topics/cut}} in order to discretize the data. Each questionnaire variable was cut to two groups, low (0) and high (1), to produce two groups as equal size as possible in length.  

We used multiple binary logistic regression to explore the relationships between all gathered variables, where one outcome variable is described by several other variables. Due to the exploratory nature of our study, we produced several models, where each of the gathered variables are used as an outcome variable. The produced models can be found in Tables ~\ref{tab:commits} to ~\ref{tab:ineffective}. 

Aikaike's information criterion (AIC) introduced by Hirotogu Akaike~\cite{akaike1973information} is an estimator of quality used to quantify the relative information lost in a model related to the data generating the model. Instead of computing p-values and significance, "the method lets you determine which model is more likely to be correct and quantify how much more likely" ~\cite{motulsky2004fitting}. AIC is widely used for model selection, as it allows to compare different models made on the same dataset. Burnham and Anderson ~\cite{burnham2004multimodel} discuss the AIC and the theoretical basis of at length, concluding that the aim is to model the information in the data rather than to model the data, and that a good model allows for separation between information and noise. 

To explore the performance of the produced models, we ran a 10-fold cross-validation, where the dataset is divided into ten different folds. We computed the area under the ROC-curve measure~\cite{bradley1997use}, abbreviated as AUC, as a performance measure for all the produced models. ROC-curve is a representation of true positive and false positive rates of the model with different threshold settings, i.e. a coin flip would produce an AUC of 0.5 and a perfect model explaining the outcome an AUC of 1.

For all the models, we checked for multicollinearity between predictors with the \emph{vif} function from the R package \emph{car}~\cite{fox2012package}. We noticed significant correlation between the number of commits and the number of lines changed, but not for the variables gathered with the questionnaire. This means in principle that models having these two variables should be removed when trying to make as accurate model as possible, but due to the exploratory nature of our models we show them in our results.
\section{Results}\label{sec:results}

\subsection*{RQ1 - \RQOne}

\begin{table}[]
\centering
\caption{Inter-coder agreement or the inter-rater reliability}
\label{tab:agreement}
\begin{tabular}{ll}
\textbf{Variable:}    		& \textbf{Krippendorff's Alpha:} \\
Hurry                            & -0.0993              \\
Independence                     & -0.119               \\
Ineffective Software Development & -0.178               \\
Interruptions                    & -0.161               \\
Sleeping Problems                & -0.214               \\
Stress                           & -0.155              
\end{tabular}
\end{table}

\textbf{Motivation}: 
Our main motivation for this research question was to understand how stressors were felt in the software project under study. In particular, if the individuals in the development team reported similar affective states at the same time during the development project. It could be that external demands such as deadlines could affect the whole development team. Answering this question allows for further processing and analysis of the data and for us to determine whether personalizing the metrics is needed or if we can build a general model pooling the whole organization's data.

Related work in investigating time pressure has called for organizational -level studies and variables free of individual biases~\cite{silla2014shared}. Additionally, there is evidence that part of work related stress is shared within organizations ~\cite{semmer1996shared}.

\textbf{Methods}: We measured how the raters agreed with each other on their daily ranking of questionnaire variables by computing Krippendorff's alpha~\cite{krippendorff2011computing}. While most studies leverage Cohen's Kappa to determine rater agreement, Krippendorff's alpha is more robust with missing data and multiple raters. The values produced by Krippendorff's alpha~\cite{krippendorff2011computing} are between 1 (perfect agreement), 0 (units statistically unrelated) and -1 (perfect disagreement). To interpret the values, Krippendorff proposed thresholds~\cite{krippendorff1980reliability}, where a value of 0.2 is considered poor, and values greater than 0.7 good.

\textbf{Findings}: \textbf{We observe poor disagreement between respondents for all questionnaire variables.} Table~\ref{tab:agreement} shows values from -0.214 for Sleeping Problems up to -0.099 for Hurry. These negative values could imply two things: either the respondents feel each affective state individually rather than on a group level, or they use different calibrations of the Likert scales. The latter implication means that some respondents might consider a value of 2 for Hurry normal while others consider it to be exceptional.

\textbf{Significance}:
As the different affective states are experienced in an individual manner and as individuals might use a different calibration of the Likert-scale, we decided to discretize each respondent's Likert-scale answers into two groups, i.e., high and low. However, instead of using a fixed threshold for each affective state, we calibrated our discretization approach by using a different threshold for each respondent, aiming for as even distribution between the groups for each individual respondent.
For a given respondent and affective state, the threshold is obtained by using standard equal probability discretization ~cite~\cite{witten2016data}. This respondent-aware discretization is done for the questionnaire variables, as well as software repository variables when they are used as outcome variables. I.e. in RQ2 we are trying to predict a binary value of low or high amount of productivity or social interaction, where as in RQ3 we transformed software repository variables to the logarithmic scale because of very large variance in the values.

\subsection*{RQ2 - \RQTwo}

\textbf{Motivation}: Prior work suggests that self-assessed affective states of valence and dominance by software developers correlate with productivity~\cite{graziotin2013happy}. For example in software engineering, time pressure experienced by developers, from both an individual~\cite{mantyla2013more, salman2018effect} and organizations~\cite{nan2009impact} (\textbf{hurry}) point of view, elevates arousal and in general increases efficiency.
Similarly, the connection between \textbf{stress} and productivity is well established~\cite{kompier1999preventing}, with stress interpreted as decreasing productivity, in extreme cases even leading to emotional exhaustion and burnout~\cite{donald2005work}. Recently, \textbf{sleeping problems} or sleep deprivation have been shown to considerably reduce implementation quality~\cite{fucci2018need} in software development. \textbf{Interruptions} to work have also been shown to increase time pressure, spent effort, felt frustration and stress~\cite{mark2008cost} in software engineering context. In the field of applied psychology the role of \textbf{Independence} (autonomy) has been under study for a mediating role between job demands and work strain perceived by employees and evidence has been found to support this notion ~\cite{bakker2005job,xanthopoulou2007job}. Finally, the impact of \textbf{inefficient software development} in terms of software processes~\cite{diaz1997software}, tools~\cite{bruckhaus1996impact} and communication~\cite{wagner2018systematic} on productivity is well documented.

The novelty of our work is that we link individual developers self-reported affective states to more objective variables acquired through repository mining. This research question aims to empirically validate the link between the above affective states and metrics of technical productivity and social interaction in the context of a software development company. 

\textbf{Metrics}: As explained in Section~\ref{sec:methodology}, we collected daily questionnaire responses for the six questionnaire variables. To measure technical productivity, we used the number of commits per day as well as the number of lines changed (i.e., churn) per day. Social interaction was measured in terms of the number of chat messages exchanged with colleagues each day. Seasonality was taken into account by introducing binary variables of all the weekdays to the equation. The variables added to control seasonality are not shown in the results as they are only added to control for spurious statistical significance that would otherwise surface.

\textbf{Method}: In this RQ, we use questionnaire answers provided in the morning about a respondent's affective state to predict the productivity and social interaction of that respondent during the rest of the day. In other words, this RQ performs within-day prediction (as opposed to RQ3).

We used a repeated 10-fold cross validation of logistic regression to model the gathered data. As indicated in RQ1, the outcome and predictor variables were discretized using person-specific thresholds to a binary variable with a scale of 0 (low) or 1 (high).

Individuals affective state does not start from the same point each morning, rather the previous days state is highly predictive of one state the next day. This is not true for measures of productivity and social interaction. The number of commits and chat messages are zero before an individual makes a commit or a chat message. Hence, we did not take autocorrelation into account for the models presented in tables~\ref{tab:commits}, \ref{tab:chat} and \ref{tab:loc}.

\textbf{Results}: In our model \textbf{decreases in technical productivity are explained by increases in sleeping problems and hurry, but decreases in stress.}
Table~\ref{tab:commits} shows the model with the discretized number of commits as outcome. We find that both self-reported sleeping problems (p-value < 0.01) and hurry (p-value < 0.001) are negatively related to the number of commits made later during the same day. Additionally, we find that self-reported stress is positively related to commit activity. For the model with an outcome value of discretized number of commits we observe an AIC value of 593.73 and an area under the curve of 0.71.

In our model \textbf{decreases in social interaction are explained by increases in sleeping problems and decreases in independence (autonomy).}
The model with the discretized number of chat messages as outcome can be found in the Table~\ref{tab:chat}.  We observe that self-reported independence is positively related to the number of chat messages, with p-values below 0.001 and a positive z-value. Additionally, sleeping problems are negatively related to the number of chat messages(p-value < 0.001). The model has an area under the curve of 0.77, performing the best out of models with an outcome variable acquired from a software repository, meaning it can explain more variance in the dataset than the models with productivity as an outcome variable.

We can find significant relationships in the model with the outcome variable of discretized lines of code, shown in the table ~\ref{tab:loc}. Self-reported sleeping problems and hurry are negatively related to lines of code changed with p-values below 0.01 in our data. Additionally, we observe that both self-reported independence and stress are positively related to the number of lines changed. The model has an AUC value of 0.72.

\textbf{Significance / Discussion}:
Previous work showed that the amount of communication between organizations~\cite{van2016speedy} tends to decrease closer towards a deadline. Likewise, employees in the information technology sector have been observed to be less reluctant to report bad news when facing time pressure~\cite{park2008overcoming}.

While our models did not find a significant relation between time pressure (\textbf{hurry}) and social interaction, they did highlight strong relations between social interaction and both \textbf{sleeping problems} and autonomy (\textbf{independence}). While the relation with sleeping problems could be due to physiological reasons, the positive relation with autonomy can imply that developers who feel being without control tend to communicate less and work in their own silo. 

On the other hand, we did find a relation of \textbf{hurry}, \textbf{stress} and (again) \textbf{sleeping problems} with technical productivity. First of all, lack of sleep is not only associated with decrease in social interaction, but also with less productivity overall. Second, while feeling more time pressure (\textbf{hurry}) is related with drops in productivity, ironically higher stress is related with increased productivity. Job demands resources model~\cite{bakker2007job} assumes that stress is the result between job demands and resources, where time pressure is part of job demands.

\begin{table}[t]
\centering
\caption{Logistic regression model for technical productivity (number of commits), controlled for weekly seasonality. Significant p-values ($\alpha=0.05$) are shown in bold.}
\label{tab:commits}
\begin{tabular}{lllll}
\textbf{Accuracy measures} & & & & \\
AIC: 593.73 & AUC: 0.71 & F1: 0.5 & Prc: 0.73 & R: 0.44 \\
\hline
\textbf{Variable} & Est & Std. E & z value & Pr(>|z|) \\
\hline
stress & 0.87 & 0.32 & 2.75 & \textbf{0.0059} \\
sleep & -0.89 & 0.26 & -3.37 & \textbf{0.00075} \\
hurry & -1.56 & 0.26 & -6.19 & \textbf{5.83e-10} \\
interruptions & -0.036 & 0.31 & -0.12 & 0.91 \\
ineffective & 0.33 & 0.33 & 0.99 & 0.32 \\
independence & -0.047 & 0.22 & -0.21 & 0.83 \\
\end{tabular}
\end{table}

\begin{table}[t]
\centering
\caption{Logistic regression model for social interaction (number of chat messages), controlled for weekly seasonality. Significant p-values ($\alpha=0.05$) are shown in bold.}
\label{tab:chat}

\begin{tabular}{lllll}
\textbf{Accuracy measures} & & & & \\
AIC: 520.67 & AUC: 0.77 & F1: 0.5 & Prc: 0.73 & R: 0.44 \\
\hline
\textbf{Variable} & Est & Std. E & z value & Pr(>|z|) \\
\hline
stress & 0.01 & 0.33 & 0.03 & 0.97 \\
sleep & -0.61 & 0.26 & -2.31 & \textbf{0.021} \\
hurry & -0.21 & 0.26 & -0.82 & 0.41 \\
interruptions & -0.50 & 0.32 & -1.56 & 0.12 \\
ineffective & 0.46 & 0.35 & 1.33 & 0.18 \\
independence & 1.9 & 0.27 & 7.03 & \textbf{2.093e-12} \\
\end{tabular}
\end{table}

\begin{table}[t]
\centering
\caption{Logistic regression model for technical productivity (number of lines changed), controlled for weekly seasonality. Significant p-values ($\alpha=0.05$) are shown in bold.}
\label{tab:loc}
\begin{tabular}{lllll}
\textbf{Accuracy measures} & & & & \\
AIC: 581.77 & AUC: 0.72 & F1: 0.52 & Prc: 0.71 & R: 0.42 \\
\hline
\textbf{Variable} & Est & Std. E & z value & Pr(>|z|) \\
\hline
stress & 0.79 & 0.32 & 2.46 & \textbf{0.014} \\
sleep & -0.74 & 0.27 & -2.76 & \textbf{0.0058} \\
hurry & -1.47 & 0.26 & -5.72 & \textbf{1.065e-08} \\
interruptions & 0.094 & 0.31 & 0.3 & 0.76 \\
ineffective & 0.48 & 0.34 & 1.41 & 0.16 \\
independence & 0.56 & 0.22 & 2.50 & \textbf{0.012} \\
\end{tabular}
\end{table}

\subsection*{RQ3 - \RQThree}

\textbf{Motivation}:
We also wanted to investigate if productivity measures predict affective states the next day. This is in opposition to RQ2, where we investigated the relationships between questionnaire variables and software repository variables during the same day. Predicting affective states from metrics obtained from software repositories could possibly be used to optimize both the productivity and well-being of software developers. In the long run, there is the possibility of replacing work satisfaction surveys with information obtained from software repositories. Another possible usage of predicting cognitive states, producing actionable and current information to project managers on the state of project workers to avoid burnouts.

\textbf{Metrics}: Here we produce models with outcome variables from the questionnaire and try to predict them with different combinations of variables obtained from the software repositories from the previous day. Instead of discretizing the raw metrics acquired from software repositories, here we transform the values into a logarithmic scale because of their highly skewed nature. To take the autocorrelation of the questionnaire variables into account, we added the previous day's response to the questionnaire to the model as a predictive variable. Similar to the models produced for RQ2, we also added binary variables of weekdays to the model to control for weekly seasonality in the data. As in RQ2 the variables added to control for the seasonality are not shown in the results. Significance of the autocorrelation for questionnaire variables can be found for models combining all software repository variables in Table ~\ref{tab:pvalues}.

\begin{table*}[t]
\centering
\caption{Models for hurry controlled for weekly seasonality and autocorrelation. Five different models shown in columns. Significant p-values ($\alpha=0.05$) are shown in bold.}
\label{tab:hurry}
\begin{tabular}{llllll}
 & {log(ncommits)
{\textsubscript{T-1}}}
 & {log(nloc){\textsubscript{T-1}}} & {log(nchat){\textsubscript{T-1}}} & {log(ncommits){\textsubscript{T-1}}} + {log(nchat){\textsubscript{T-1}}} & {log(ncommits)
{\textsubscript{T-1}}}
 + {log(nloc){\textsubscript{T-1}}} + {log(nchat){\textsubscript{T-1}}} \\ 
 \hline
z-value                                     & -3.528             & -2.713     & -2.384                 & -3.067  \& 1.571                                                                                                                                                           & -2.350  \& 1.056  \& -1.487    \\
p-value                                     & \textbf{0.000418}        & \textbf{0.00667}  & \textbf{0.00667}                 & \textbf{0.0171} \& \textbf{0.00216}                                                                                                                                                          & \textbf{0.0188} \& 0.2911 \& 0.1369 \\
AIC                                         & 157.25          & 163 & 164.41                 & 156.77                                                                                                                                                                     & 157.69                     \\
10-fold AUC                                 & 0.906          & 0.884     & 0.906                 & 0.919                                                                                                                                                                     & 0.917                    \\
Precision                                   & 0.926           & 0.930      & 0.930            & 0.922&       0.923                     \\
Recall                                      & 0.951          & 0.947     & 0.947                  & 0.949                                                                                                                                                                      & 0.949                      \\
F1 Score                                    & 0.937           & 0.938      & 0.938                  & 0.934                                                                                                                                                                     & 0.934                     
\end{tabular}
\end{table*}

\begin{table*}[t]
\caption{~Models for sleeping problems controlled for weekly seasonality and autocorrelation. Five different models shown in columns. Significant p-values ($\alpha=0.05$) are shown in bold.}
\label{tab:sleep}
\begin{tabular}{llllll}
      & {log(ncommits)
{\textsubscript{T-1}}}
 & {log(nloc){\textsubscript{T-1}}} & {log(nchat){\textsubscript{T-1}}} & {log(nloc){\textsubscript{T-1}}} + {log(nchat){\textsubscript{T-1}}} & {log(ncommits)
{\textsubscript{T-1}}}
 + {log(nloc){\textsubscript{T-1}}} + {log(nchat){\textsubscript{T-1}}} \\
 \hline
z-value     & -2.345       & -2.807    & -2.249     & -2.397 \& -1.688       & 0.752 \& -1.659 \& -1.774               \\
p-value     & \textbf{0.019}        & \textbf{0.005}     & \textbf{0.0245}     & \textbf{0.0165} \& 0.0915       & 0.4519 \& 0.0971 \& 0.0761              \\
AIC         & 158.66        & 156.16    & 159.19     & 155.35                 & 156.77                                 \\
10-fold AUC & 0.869         & 0.872    & 0.873     & 0.873                 & 0.863                                 \\
Precision   &     0.926         &    0.927       &   0.923
&    0.931       & 0.933                                  \\
Recall      &    0.939    &  0.941      &   0.931  &   0.941    & 0.944                                 \\
F1 Score    &   0.931  &    0.933      &   0.926  &   0.935   & 0.938                                 
\end{tabular}
\end{table*}

\begin{table*}[!t]
\caption{Models for stress controlled for weekly seasonality and autocorrelation. Five different models shown in columns. Significant p-values ($\alpha=0.05$) are shown in bold.}
\label{tab:stress}
\begin{tabular}{llllll}

      & {log(ncommits)
{\textsubscript{T-1}}}
 & {log(nloc){\textsubscript{T-1}}} & {log(nchat){\textsubscript{T-1}}} & {log(nloc){\textsubscript{T-1}}} + {log(nchat){\textsubscript{T-1}}} & {log(ncommits)
{\textsubscript{T-1}}}
 + {log(nloc){\textsubscript{T-1}}} + {log(nchat){\textsubscript{T-1}}} \\
 \hline
 z-value     & -0.080        & 0.483   & -2.524     & 1.275 \& -2.772        & -0.344 \& 0.981 \& -2.627       \\
p-value     & 0.936        & 0.629       & \textbf{0.016}     & 0.20241 \& \textbf{0.00558}     & 0.731 \& 0.326 \& \textbf{0.00862}             \\
AIC         & 209.43       & 209.2     & 202.95     & 203.3                 & 205.18                                 \\
10-fold AUC & 0.805       & 0.82     & 0.844     & 0.845   &0.841                                       \\
Precision   & 0.892       & 0.899     & 0.90      & 0.893     & 0.893    \\
Recall      & 0.887        & 0.887      & 0.898      & 0.896   & 0.896                  \\
F1 Score    & 0.892        & 0.891      & 0.897      & 0.893                  & 0.893                                 
\end{tabular}
\end{table*}

\textbf{Results}: For each outcome variable acquired by the questionnaire, we present five different models for each outcome variable from the questionnaire in Tables ~\ref{tab:hurry} to ~\ref{tab:ineffective}. To explore the relationships with all software repository variables, we show five different models: one model with each of software repository variables as the predictor, one model with the lowest AIC of all the possible combinations of two software repository variables, and lastly the combination of all three software repository variables as predictors. Five different models are shown in the columns, while coefficients and accuracy measures for the models can be seen in different rows. 

\textbf{We find that the number of chat messages are negatively related to hindered work well-being(stress, hurry, sleeping problems and felt ineffective software development). Additionally, we observe a positive relationship with self-reported independence}. All significant p-values with all three software repository metrics as predictors are displayed in Table ~\ref{tab:pvalues}. 

The links between productivity variables and questionnaire variables are not as strong as with a measure of social interaction. However, \textbf{We find a significant negative relationship between productivity measures and self-reported hurry and sleeping problems the next day.} The effect of this relationship decreases when multiple software repository variables are introduced as predictors, but it can be found most clearly when commits or lines code are used solely as predictors. This is shown in Tables ~\ref{tab:hurry} and ~\ref{tab:sleep}. Additionally, when models are made with two software repository variables, we can find a positive significant relationship between self-reported interruptions and number of lines of code changed, as can be seen in Table ~\ref{tab:interruptions}. 

We generally observe very high area under the roc-curve for the produced models, between 76\% to over 95\% for models with independence as outcome variable. We believe this is due to very strong autocorrelations in the questionnaire variables, which are highlighted in table ~\ref{tab:pvalues}. 

\begin{table*}[!t]
\caption{Models for interruptions controlled for weekly seasonality and autocorrelation. Five different models shown in columns. Significant p-values ($\alpha=0.05$) are shown in bold.}
\label{tab:interruptions}
\begin{tabular}{llllll}
      & {log(ncommits)
{\textsubscript{T-1}}}
 & {log(nloc){\textsubscript{T-1}}} & {log(nchat){\textsubscript{T-1}}} & {log(nloc){\textsubscript{T-1}}} + {log(nchat){\textsubscript{T-1}}} & {log(ncommits)
{\textsubscript{T-1}}}
 + {log(nloc){\textsubscript{T-1}}} + {log(nchat){\textsubscript{T-1}}} \\
 \hline
z-value     & 0.926        & 1.684      & -1.964     & 2.322 \& -2.562        & -0.414 \& 1.697 \& -2.397              \\
p-value     & 0.335        & 0.0922      & \textbf{0.049}     & \textbf{0.0202} \& \textbf{0.0104}     & 0.6789 \& 0.0897 \& \textbf{0.0165}          \\
AIC         & 225.09       & 222.97     & 222.15     & 218.36                 & 220.19                                 \\
10-fold AUC & 0.805       & 0.821     & 0.815     & 0.835                 & 0.829                                 \\
Precision   & 0.881        & 0.879      & 0.877     & 0.863                   & 0.861                                 \\
Recall      & 0.881        & 0.881     & 0.891      & 0.885                  & 0.885                                  \\
F1 Score    & 0.88        & 0.878      & 0.882      & 0.872                   & 0.871                                 
\end{tabular}
\end{table*}

\begin{table*}[!t]
\centering
\caption{Model for independence controlled for weekly seasonality and autocorrelation. Five different models shown in columns. Significant p-values ($\alpha=0.05$) are shown in bold.}
\label{tab:independence}
\begin{tabular}{llllll}
      & {log(ncommits)
{\textsubscript{T-1}}}
 & {log(nloc){\textsubscript{T-1}}} & {log(nchat){\textsubscript{T-1}}} & {log(ncommits){\textsubscript{T-1}}} + {log(nchat){\textsubscript{T-1}}} & {log(ncommits)
{\textsubscript{T-1}}}
 + {log(nloc){\textsubscript{T-1}}} + {log(nchat){\textsubscript{T-1}}} \\
 \hline
z-value     & 0.718        & -0.060      & 2.436      & 0.088 \& 2.318             & 0.791 \& -0.89 \& 2.152               \\
p-value     & 0.473        & 0.952       & \textbf{0.0149}      & 0.9302 \& \textbf{0.0205}           & 0.4290 \& 0.3728 \& \textbf{0.0314}              \\
AIC         & 120.5       & 121.06     & 115.16     & 117.16                     & 118.37                                                                      \\
10-fold AUC & 0.949       & 0.938      & 0.956     & 0.955      & 0.949       \\
Precision   & 0.958        & 0.958      & 0.958      & 0.958                      & 0.958                                  \\
Recall      & 0.942         & 0.942       & 0.942      & 0.942                     & 0.940                                  \\
F1 Score    & 0.949        & 0.949      & 0.949      & 0.949                      & 0.948                                 
\end{tabular}
\end{table*}

\begin{table*}[!t]
\centering
\caption{Models for ineffective software development controlled for weekly seasonality and autocorrelation. Five different models shown in columns. Significant p-values ($\alpha=0.05$) are shown in bold.}
\label{tab:ineffective}
\begin{tabular}{llllll}
      & {log(ncommits)
{\textsubscript{T-1}}}
 & {log(nloc){\textsubscript{T-1}}} & {log(nchat){\textsubscript{T-1}}} & {log(nloc){\textsubscript{T-1}}} + {log(nchat){\textsubscript{T-1}}} & {log(ncommits)
{\textsubscript{T-1}}}
 + {log(nloc){\textsubscript{T-1}}} + {log(nchat){\textsubscript{T-1}}} \\
 \hline
z-value     & 0.510        & 1.286      & -1.847     & 1.745 \& -2.205        & -0.309 \& 1.365 \& -2.062               \\
p-value     & 0.610        & 0.199      & 0.0647    & 0.0810 \& \textbf{0.0275}   & 0.757 \& 0.1724\& \textbf{0.0392}       \\
AIC         & 150.76        & 149.21     & 147.66     & 146.31                 & 148.21                                 \\
10-fold AUC & 0.757       & 0.798     & 0.788     & 0.819                 & 0.81                                 \\
Precision   & 0.936        & 0.935      & 0.935      & 0.925                  & 0.925                                  \\
Recall      & 0.948        & 0.95      & 0.95      & 0.952                  & 0.952                                 \\
F1 Score    & 0.941        & 0.941      & 0.942      & 0.937                 & 0.937                                 
\end{tabular}
\end{table*}

\begin{table}[]
\centering
\caption{Significant p-values of models with the outcome variable of questionnaire variable and all software repository variables as predicting variables.}
\label{tab:pvalues}
\begin{tabular}{lllllll}
\textbf{Model/Variable} & ncommits & nloc & nchat & autocorrelation \\
\hline
stress &  &  & 0.00862 & 9.02e-16 \\
sleep &  &  &  & 1.14e-14 \\
hurry &  & 0.0188 &  & 2.6e-15 \\
interruptions &  &  & 0.0165 & 4.08e-13 \\
ineffective &  &  & 0.0314 & <2e-16 \\
independence &  &  & 0.0392 & 5.24e-11 \\
\end{tabular}
\end{table}

\textbf{Significance / Discussion}:
Yerkes-Dodson's law~\cite{yerkes1908relation} dictates that there is a U-shaped relationship between arousal and performance, i.e., time pressure increases performance up unto a certain point before it starts decreasing again. Our results are partly supported by Yerkes-Dodson's law, as developers who reported experiencing hurry had less commits and lines of code changed the previous day, i.e., the second part of the Yerkes-Dodson law. This could in part be explained by the self-reported nature of the variable, since developers who were more productive and communicative the previous day simply could have felt less hurry the next day when they answered the questionnaire.

Additionally, different kinds of tasks are performed during different phases of a software project, which could have an effect, e.g., more commits and lines of code are changed at the start of developing a feature, whereas work finishing a feature before a deadline could change relatively little. Having less communication before a deadline has been observed between organizations in prior work~\cite{van2016speedy}, we find that developers who felt more hurry also had less social interaction with their own team prior to answering the questionnaire.

We also observe that developers who reported having sleeping problems were less productive and had less social interaction during the previous day. This result is in line with results from various fields such as applied psychology, where it has been observed that performance in complex cognitive tasks decreases both with consecutive hours of wakefulness as well as consecutive days of disrupted sleep ~\cite{wickens2015impact}. Similar results have been recently acquired by Fucci et al~\cite{fucci2018need}, where the quality of work lessened with sleep deprivation.

We cannot observe a link between simple productivity measures obtained from the software repositories and self-reported stress. Such a link could be expected to be found based on prior work in other fields~\cite{kompier1999preventing}. In previous work, Miller et al.~\cite{miller1990integrated} noticed that perception of participation in the decision making process reduces role stress. In other words, when there is communication about the distribution of work tasks stress is lessened. In line with these results, we observe a negative link between number of chat messages and self-reported stress. It must also be noted that occupational differences in the relationship between the stress and communication were reported previously~\cite{miller1989occupational}.

There is a negative link between self-reported level of interruptions and the amount of chat messages posted in the data we have gathered. This is partly in conflict with previous findings, for example Cameron and Webster~\cite{cameron2005unintended} report that the interruptive nature of instant messaging is considered unfair by employees. Additionally, the prohibition of instant messaging in software development companies has also been reported ~\cite{sykes2011interruptions}. It must be noted that the way in which instant messaging is used affects the results, i.e., whether instant messaging is used for coordination and social bonding between the development team or to interrupt work by forwarding new commands to developers. Investigating the content and quality of these communication channels is needed for further conclusions.

Respondents who reported having the ability to make independent decisions on their work had more chat messages on the previous day. Our results are partly in line with the model by Karasek~\cite{karasek1979job}, which proposes that worker autonomy increases worker well-being. While our models cannot link self-reported independence to productivity measures, as could be expected from previous work~\cite{bakker2005job,xanthopoulou2007job}, the relationship between chat messages and perceived independence is positive, while in the other models the relationship between chat messages and hurry, sleeping problems, stress and interruptions is negative.

We can observe a negative link between self-reported ineffective software development and the number of chat messages. As self-reported ineffective software development is in part defined as poor communication this result makes sense. In prior work, it has been noted that coordinating expertise in software teams is associated with stronger team performance~\cite{faraj2000coordinating}.
\section{Threats to validity}\label{sec:future}

The questionnaire was taken only in a single software company with a single software project. This diminishes the generalizability of our results. To achieve higher response rates, we opted for six single items on the questionnaire. Because of this, we cannot estimate the internal validity of the questionnaire. However, variety of seminal studies have shown that single items on questionnaires produce valid data ~\cite{wanous1997overall, elo2003validity, nagy2002using}.

One limitation of the analysis is the form of the data. As the questionnaire is only taken during working days, and no questionnaire answers are given during the weekend, as a result around 20\% of the responses do not count for the regression analysis at all and the models with questionnaire variables as outcomes in Tables~\ref{tab:sleep}, \ref{tab:stress}, \ref{tab:hurry}, \ref{tab:interruptions}, \ref{tab:independence} and \ref{tab:ineffective}. Additionally, responses without answers from adjacent days are not taken into account when building some of the models. This is because autocorrelation is taken into account with a lagged variable, and as such predictions with the model can only be made for around half of the dataset. Affective states are effected also by events and experiences from private lives that are not at all work related.

Our analysis only investigates easily quantifiable metrics from information systems, but it does not investigate the quality behind these metrics such as the quality of the commits, or the topics of the chat messages. In the future, in addition to gathering data containing the quality of work and emotional sentiment of the communication, we want to study the relationship of events such as meetings and build breakages to the self-reported variables acquired with the questionnaire answers. 
\section{Conclusions}\label{sec:conclusions}

In this paper, we investigated software developers self-reported well-being with an experience sampling method and linked it to metrics mined from software repositories in the context of a single software project. We found no shared emotions from the survey, i.e., for none of the questionnaire variables we observed agreement between the respondents. Thus we opted for personal discretization of the variables.

We make three important findings and contribution. First, when developers reported high hurry they were surprisingly less productive, i.e. made less commits, and had less lines of code changed. 
Hence our study captures the opposite side of Yerkes-Dodson law \cite{yerkes1908relation} than previous experiments in software engineering \cite{mantyla2014time,mantyla2013more,salman2018effect}.

Second, we find that all variables related to hindered work well-being (hurry, stress, sleeping problems, interruptions and ineffective software development) are negatively related to the number of chat messages posted during the previous day. On the contrary, developers with high self-reported feeling of independence (autonomy) chat more both during the same and the previous day.

Third, this paper presents the first study in software engineering using experience sampling method and connecting it with software repositories. We investigate experienced emotions and work health with experience sampling. However, we think that many other factors in the domain of software engineering could be studied with this method. For example, investigating the adaption of Agile practices or process changes would be an interesting candidate for experience sampling study, as day to day data from practice and process changes could be observed and investigated. 

\begin{acks}
The first, second and third author have been supported by Academy of Finland grant 298020. The first author has been supported by Kaute-foundation. 
\end{acks}
\bibliographystyle{ACM-Reference-Format}
\bibliography{sample-bibliography}

\end{document}